# Improved code-based identification scheme

Pierre-Louis Cayrel
CASED
Mornewegstrasse, 32
64293 Darmstadt
Germany
pierre-louis.cayrel@cased.de

Pascal Véron
IMATH
Université de Toulon et du Var.
B.P. 20132, F-83957 La Garde Cedex,
France
veron@univ-tln.fr

*Abstract*—We revisit the 3-pass code-based identification scheme proposed by Stern at Crypto'93, and give a new 5-pass protocol for which the probability of the cheater is $\approx 1/2$ (instead of 2/3 in the original Stern's proposal). Furthermore, we propose to use quasi-cyclic construction in order to dramatically reduce the size of the public key.

The proposed scheme is zero-knowledge and relies on an NP-complete problem coming from coding theory (namely the $q$-ary Syndrome Decoding problem). Taking into account a recent study of a generalization of Stern's information-set-decoding algorithm for decoding linear codes over arbitrary finite fields $\mathbb{F}_q$, we suggest parameters so that the public key be 34Kbits while those of Stern's scheme is about 66Kbits. This provides a very practical identification (and possibly signature) scheme which is mostly attractive for light-weight cryptography.

## I. INTRODUCTION

Cryptosystem based on number-theory (problems of factorisation and discrete logarithm) is more and more widely used in the real world. After Shor's algorithm which describes a quantum algorithm to solve in polynomial time the two previous problems, there is a strong need for public key schemes which are not based on such problems. Firstly because it would be unreasonable to consider only one type of hard problem. At the time, nearly all public key cryptographic products are based on integer factorization or discrete logarithm. Secondly, even if the above mentioned problems remain hard, practical progress in factorization and discrete logarithm computation leads to choose larger and larger keys.

The Shor's algorithm doesn't threaten the so-called *post-quantum* cryptosystems as lattice-based, code-based, and multivariate-based cryptosystems.

In this paper, we consider a particular type of alternative cryptography, based on error-correcting code theory. Code-based cryptography was initiated a long time ago with the celebrated McEliece encryption algorithm.

Public key identification (ID) protocols allow a party holding a secret key to prove its identity to any other entity holding the corresponding public key. The minimum security of such protocols should be that a passive observer who sees the interaction should not then be able to perform his own interaction and successfully impersonate the prover.

At Crypto'93, Stern proposed a new scheme, which is still today the reference in this area [18]. The Stern's scheme is a multiple round zero-knowledge protocol, where each round is a three-pass interaction between the prover and the verifier. Stern' scheme has two major drawbacks :

1) many rounds are required because the cheater success probability is 2/3 instead of 1/2 for Fiat-Shamir's factorization-based protocol [8], hence typically 27 rounds are needed so that this probability be less than $2^{-16}$,
2) the public key is very large, typically 66 Kbits.

The first issue was addressed by Gaborit and Girault in [10] and the second one was partially adressed by Véron [20]. In this paper, we focus on the first drawback. Using $q$-ary codes instead of binary ones, we define a 5 pass identification scheme for which the probability of a cheater is bounded by 1/2. We then propose to use quasi-cyclic construction to address the second drawback.

*Organisation of the paper*

In Section II, we give basics facts about code-based cryptography, we describe the original Stern's scheme and propose in Section III a new identification scheme which permits to reduce the number of rounds involved during the identification process. In Section IV, we describe the properties of our proposal and study its security. The Section V concludes our contribution.

## II. CODE-BASED CRYPTOGRAPHY

In this section we recall basic facts about code-based cryptography. We refer to [4], for a general introduction to these problems.

### A. Definitions

Linear codes are $k$-dimensional subspaces of an $n$-dimensional vector space over a finite field $\mathbb{F}_q$, where $k$ and $n$ are positive integers with $k < n$, and $q$ is a prime power. The error-correcting capability of such a code is the maximum number $t$ of errors that the code is able to decode. In short, linear codes with these parameters are denoted $(n, k, t)$-codes.

**Definition II.1 (Hamming weight)** *The (Hamming) weight of a vector $x$ is the number of non-zero entries. We use $\mathsf{wt}(x)$ to represent the Hamming weight of $x$.*

**Definition II.2 (Generator and Parity Check Matrix)** *Let $\mathscr{C}$ be a linear code over $\mathbb{F}_q$. A generator matrix G of $\mathscr{C}$ is a matrix whose rows form a basis of $\mathscr{C}$:*

$$\mathscr{C} = \{xG : x \in \mathbb{F}_q^k\}.$$

*A parity check matrix H of $\mathscr{C}$ is defined by*

$$\mathscr{C} = \{x \in \mathbb{F}_q^n : Hx^T = 0\}$$

*and generates the dual space of $\mathscr{C}$.*

We describe here the main hard problems on which are based the code-based cryptosystems.

Let $n$ and $r$ be two integers such that $n \geq r$, $\mathsf{Binary}(n,r)$ (resp. $\mathsf{q-ary}(n,r)$) be the set of binary (resp. $q$-ary) matrices with $n$ columns and $r$ rows of rank $r$. Moreover, let us denote by $x \xleftarrow{R} A$, the fact that $x$ is randomly selected in the set $A$.

**Definition II.3 (binary Syndrome Decoding (SD) problem)**

Input : $H \xleftarrow{R} \mathsf{Binary}(n,r)$, $y \xleftarrow{R} \mathbb{F}_2^r$ and an integer $\omega > 0$.
Ouput : *A word $s \in \mathbb{F}_2^n$ such that $\mathsf{wt}(s) \leq \omega$, $Hs^T = y$.*

This problem was proven to be NP-complete in 1978 [3], but only for binary codes.

**Definition II.4 ($q$-ary Syndrome Decoding ($q$SD) problem)**

Input : $H \xleftarrow{R} \mathsf{q-ary}(n,r)$, $y \xleftarrow{R} \mathbb{F}_q^r$ and an integer $\omega > 0$.
Ouput : *A word $s \in \mathbb{F}_q^n$ such that $\mathsf{wt}(s) \leq \omega$, $Hs^T = y$.*

In 1994, A. Barg proved that this result holds for codes over all finite fields [1, in russian].

The problems which cryptographic applications rely upon can have different numbers of solutions. For example, public key encryption schemes usually have exactly one solution, while digital signatures often have more than one possible solution. For code-based cryptosystems, the uniqueness of solutions can be expressed by the Gilbert-Varshamov (GV) bound :

**Definition II.5 ($q$-ary Gilbert-Varshamov bound)**
*Let $H_q(x)$ be the $q$-ary entropy function, given by :*

$$H_q(x) = x\dot{\log}_q(q-1) - x\dot{\log}_q x - (1-x)\dot{\log}_q(1-x).$$

*Suppose $0 \leq \delta \leq (q-1)/q$. Then there exists an infinite sequence of $(n,k,d)$ $q$-ary linear codes with $d/n = \delta$ and rate $R = k/n$ satisfying the inequality :*

$$R \geq 1 - H_q(\delta) \qquad \forall n.$$

### B. SD identification schemes

Stern's scheme is the first practical zero-knowledge identification scheme based on the Syndrome Decoding problem [18]. The scheme uses a binary $(n-k) \times n$ matrix $H$ common to all users. If $H$ is chosen randomly, it will provide a parity check matrix for a code with asymptotically good minimum distance given by the (binary) Gilbert-Varshamov (GV) bound. The private key for a user will thus be a word $s$ of low weight $\mathsf{wt}(s) = \omega$ (with $H_2(\omega) \approx 1 - k/n$), which sums up to the syndrome $Hs^T = y$, the public key. By Stern's 3-pass zero-knowledge protocol, the secret key holder can prove his knowledge of $s$ using two blending factors: a permutation and a random vector. However, a dishonest prover not knowing $s$ can cheat the verifier in the protocol with probability $2/3$. Thus, the protocol has to be run several times to detect cheating provers. The security of the scheme relies on the difficulty of the general decoding problem, that is on the difficulty of determining the preimage $s$ of $y = Hs^T$. As mentioned in [3], the SD problem, stated in terms of generator matrix is also NP-complete since one can go from the parity-check to the generator matrix (or vice-versa) in polynomial time. In [20], the author uses a generator matrix of a random linear binary as the public key and defines this way a dual version of Stern's scheme in order to obtain, among other things, an improvement of the transmission rate.

Figure 1 sums up the performances of the two 3-pass SD identification schemes for a probability of cheating bounded by $10^{-6}$. The computation complexity is the number of bits operation involved by the protocol and the communication complexity the number of exchanged bits.

|  | SD | G-SD |
|---|---|---|
| Rounds | 35 | 35 |
| Public data (bits) | 66048 | 66816 |
| Computation complexity | $2^{22.13}$ | $2^{22.5}$ |
| Communication complexity | 40133 | 34160 |

Fig. 1. Performances of SD schemes

### C. Attacks

For SD identification schemes, since the matrix used is a random one, the cryptanalyst is faced to the problem of decoding a random binary linear code. There are two main families of algorithms to solve this problem : Information Set Decoding (ISD) and (Generalized) Birthday Algorithm (GBA). The Information-Set-Decoding Attack seems to have the lowest complexity.

One tries to recover the $k$ information symbols as follows : the first step is to pick $k$ of the $n$ coordinates randomly in the hope that all of them are errors free. Try then to recover the message by solving a $k \times k$ linear system (binary or over $\mathbb{F}_q$).

In [15], the author presents a generalization of Stern's information-set-decoding algorithm from [17] for decoding linear codes over arbitrary finite fields $\mathbb{F}_q$ and analyzes the complexity. We will choose our parameters with regards to the complexity of this attack.

## III. A NEW IDENTIFICATION SCHEME

In what follows, we consider an element of $\mathbb{F}_q^n$ as $n$ blocs of size $\lceil \log_2(q) \rceil = N$. We represent each element of $\mathbb{F}_q$ as $N$ bits. We first introduce a special transformation that we will use in our protocol.

**Definition III.1** *Let $\Sigma$ be a permutation of $\{1,\ldots,n\}$ and $\gamma = (\gamma_1,\ldots,\gamma_n) \in \mathbb{F}_q^n$ such that $\forall i, \gamma_i \neq 0$. We define the transformation $\Pi_{\gamma,\Sigma}$ as :*

$$\Pi_{\gamma,\Sigma} : \begin{array}{rcl} \mathbb{F}_q^n & \longrightarrow & \mathbb{F}_q^n \\ v & \mapsto & (\gamma_{\Sigma(1)} v_{\Sigma(1)}, \ldots, \gamma_{\Sigma(n)} v_{\Sigma(n)}) \end{array}$$

Notice that $\forall \alpha \in \mathbb{F}_q, \forall v \in \mathbb{F}_q^n$,

$$\Pi_{\gamma,\Sigma}(\alpha v) = \alpha \Pi_{\gamma,\Sigma}(v) \text{ and } \mathsf{wt}(\Pi_{\gamma,\Sigma}(v)) = \mathsf{wt}(v).$$

### A. Key generation

Let $r = n - k$, the scheme uses a random $(r \times n)$ $q$-ary matrix $H$ common to all users. It can be considered as the parity check matrix of a random linear $(n,k)$ $q$-ary code. Without loss of generality, we can assume that $H$ is given under the form $H = (I_r | M)$ where $M$ is a random $r \times r$ matrix, since it is well known that a Gaussian elimination doesn't change the code generated by $H$. Let $\kappa$ be the security parameter, algorithm 1 describes the key generation process.

---

**Algorithm 1** Key generation($1^\kappa$)

▷ Choose $n, k, \omega$ and $q$ such that $\mathrm{WF}_{\mathrm{ISD}}(n,r,\omega,q) \geq 2^\kappa$
▷ Randomly pick a $(r \times n)$ $q$-ary matrix $H$.
▷ Randomly pick $s \in \mathbb{F}_q^n$ with $\mathsf{wt}(s) = \omega$.
▷ Compute $y$ such that $Hs^T = y$.
▷ Output : pk $= (H, y, \omega)$ and sk $= s$.

---

### B. Identification

The secret key holder can prove his knowledge of $s$ using two blending factors: the transformation above mentionned and a random vector. However, a dishonest prover not knowing $s$ can cheat the verifier in the protocol with probability $\approx 1/2$. Thus, the protocol has to be run several times to detect cheating provers. The security of the scheme relies on the difficulty of the general decoding problem, that is on the difficulty of determining the preimage $s$ of $y = Hs^T$.

This protocol is repeated $\delta$ times. A cheater has, once $c_1$ and $c_2$ sent, to be able to answer to $2q$ possible questions. If he is only able to answer to $q + 2$ questions then he is able to find a solution to the problem. Indeed, let us denote by $z$ the value sent when $b = 1$, for $c_1$ and $c_2$ fixed, this means that there exist $\alpha$ and $\alpha'$ distinct and $\beta$ and $\beta'$ such that :

$$H\Pi_{\gamma,\Sigma}^{-1}(\beta)^T - \alpha y = H\Pi_{\gamma,\Sigma}^{-1}(\beta')^T - \alpha' y$$
$$\beta - \alpha z = \beta' - \alpha' z$$

---

**Algorithm 2** Identification Scheme

**Private key**, sk: $s \in \mathbb{F}_q^n$ such that $Hs^T = y$ and $\mathsf{wt}(s) = \omega$.

**Public key**, pk: $H$ a $(n - k \times n)$ random matrix of rank $n - k$ over $\mathbb{F}_q$, $h$ a collision resistance hash function, $y \in \mathbb{F}_q^{n-k}$ and $\omega \in \mathbb{N}$

▷ Prover: generates a vector $u \in \mathbb{F}_q^n$, a vector $\gamma \in \mathbb{F}_{q*}^n$ and a permutation $\Sigma$ over $\{1,\ldots,n\}$ at random and computes the commitments :

1: Set $c_1 \leftarrow h\left(\Sigma, \gamma, Hu^T\right)$
2: Set $c_2 \leftarrow h\left(\Pi_{\gamma,\Sigma}(u), \Pi_{\gamma,\Sigma}(s)\right)$
3: Sends the commitments $\{c_1, c_2\}$ to the Verifier.
   ▷ Verifier: chooses a random $\alpha \in \mathbb{F}_q$ and sends it to the Prover.
   ▷ Prover: sends $\Pi_{\gamma,\Sigma}(u + \alpha s) = \beta \in \mathbb{F}_q^n$ to the Verifier.
   ▷ Verifier: sends a challenge $b \in \{0,1\}$ to the Prover.
   ▷ Prover: answers the challenge
4: **if** $b = 0$ **then** reveals $\Sigma$ and $\gamma$.
5: **else if** $b = 1$ **then** reveals $\Pi_{\gamma,\Sigma}(s)$.
6: **end if**
   ▷ Verifier: checks commitment correctness
7: **if** $b = 0$ **then** checks if $c_1 = h(\Sigma, \gamma, H\Pi_{\gamma,\Sigma}^{-1}(\beta)^T - \alpha y)$ is correct
8: **else if** $b = 1$ **then** checks if $c_2 = h(\beta - \alpha\Pi_{\gamma,\Sigma}(s), \Pi_{\gamma,\Sigma}(s))$ is correct and if $\mathsf{wt}(\Pi_{\gamma,\Sigma}(s)) = \omega$.
9: **end if**

---

with $\mathsf{wt}(z) = \omega$. Which gives $\beta - \beta' = (\alpha - \alpha')z$ and $H\Pi_{\gamma,\Sigma}^{-1}(\beta - \beta')^T = (\alpha - \alpha')y$. We then deduce $H\Pi_{\gamma,\Sigma}^{-1}(z)^T = y$ and $\mathsf{wt}(\Pi_{\gamma,\Sigma}^{-1}(z)) = \mathsf{wt}(z) = \omega$.

Hence the success probability of a cheater is bounded by $\frac{q+1}{2q}$.

### C. Signature

By using the so-called Fiat-Shamir paradigm [8], it is theoretically possible to convert this protocol into a signature scheme, even if it is practically questionable, since the signature is large.

## IV. PROPERTIES AND SECURITY OF THE SCHEME

### A. Zero-knowledge

Let $I = (H, y, \omega)$ be the public data shared by the prover and the verifier and let $P(I, s)$ be the following predicate :

$P(I, s) = $ "$s$ is a vector which satisfies $Hs^T = y, \mathsf{wt}(s) = \omega$" then :

**Proposition IV.1** *The protocol is an interactive proof of knowledge for $P(I, s)$.*

Due to the limited size of this paper we don't detail this proposition here.

**Theorem IV.1** *The protocol is a zero-knowledge interactive proof for $P(I, s)$ in the random oracle model.*

We will detail this theorem in a longer version of this article. However, notice that because of the randomness of $u$ and $\alpha$ only random values are exchanged during the protocol.

*B. Security and Parameters*

Like binaries SD identification schemes, security of our scheme relies on three properties of random linear $q$-ary codes :
1) Random linear codes satisfy the $q$-ary Gilbert-Varshamov lower bound [11];
2) For large $n$ almost all linear codes lie over the Gilbert-Varshamov bound [16];
3) Solving the $q$-ary syndrome decoding problem for random codes is NP-complete [1].

Now taking into account the bounds on the workfactor of the Information Set Decoding algorithm over $\mathbb{F}_q$ given in [14] (and which generalizes the bounds given in [9]) we have to set up parameters in order to obtain a practical scheme with a security level greater or equal than $2^{80}$. We have then to choose the number of rounds in order to minimize the probability of success of a cheater.

Since we deal with random codes, we have to select parameters with respect to the Gilbert-Varshamov bound (see Definition II.5), i.e. choose a weight $\omega$ with respect to this bound. Moreover as usual we will now suppose $k = r = n/2$. Let $N$ be the number of bits needed to encode an element of $\mathbb{F}_q$, $\ell_h$ the output size of the hash function $h$, $\ell_\Sigma$ *(resp. $\ell_\gamma$)* the size of the seed used to generate the permutation $\Sigma$ *(resp. the vector $\gamma$)* and $\delta$ the number of rounds. We have the following properties for our scheme :

Size of the public data in bits :

$$k \times k \times N + nN \text{(we use the systematic form of } H)$$

Total number of bits exchanged :

$$\delta(2\ell_h + N + nN + 1 + (\ell_\Sigma + \ell_\gamma + nN)/2)$$

Computation complexity over $\mathbb{F}_q$ :

$$\delta(k + n + 2\mathsf{wt}(s) \text{ multiplications} + k + \mathsf{wt}(s) \text{ additions})$$

To obtain a precise complexity on the workfactor of ISD algorithms over $\mathbb{F}_q$ we've used the code developped by C. Peters which estimates (using a Markov chain implementation) the number of iterations of the best algorithm for an attack using all possible known tricks [15]. ISD algorithms depend on a set of parameters and this code allows to test which ones can minimize the complexity of the attack.

We suggest to use for our scheme :

$$q = 256, n = 128, k = 64, \mathsf{wt}(s) = 49.$$

The complexity of an attack using ISD algorithms is then at least $2^{87}$. For the same security level in SD schemes, we need to take $n = 700, k = 350, \mathsf{wt}(s) = 75$. Table 2 sums up the characteristics of our scheme and those of SD schemes for a same level of security and a probability of cheating of $2^{-16}$. We considered that all seeds used are of length 128.

|  | SD | G-SD | Our scheme |
|---|---|---|---|
| Rounds | 27 | 27 | 16 |
| Public data (bits) | 123200 | 124250 | 33792 |
| Communication (bits) | 37872 | 31572 | 30848 |
| Computation | $2^{22.7}$ | $2^{23.1}$ | $2^{12.1}$mult+ |
|  | (bits. op) | (bits. op) | $2^{11.3}$add |
|  |  |  | (bytes op.) |

Fig. 2. SD schemes vs. $q$-ary SD scheme

To obtain a security level of $2^{128}$ suggested parameters are

$$q = 256, n = 208, k = 104, \mathsf{wt}(s) = 78$$

which gives a scheme with the following properties :

Rounds : 16
Public data (bits) : 88192
Communication (bits) : 46224
Computation (bytes op.) : $2^{12.9}$mult. and $2^{11.5}$add.

Notice that in [18], Stern has proposed two 5 pass variant of his scheme : one to minimize the computing load and the other one to lower the number of rounds. However, for these two variants, the size of the public data and the communication complexity are greater than the one of our scheme. A precise comparison for the computation complexity will be made in a longer version of this paper.

*C. Reducing public key size*

*1) Quasi-cyclic construction:* In [10], the authors propose a variation of the Stern identification scheme by using double circulant codes. The circulant structure of the public matrix makes the computation very easy without having to generate the whole binary matrix, indeed the whole scheme only needs very few memory storage. They propose a scheme with a public key of size 347 bits and a private key of size 694 bits.

We can use this construction in our context by replacing the random $q$-ary matrix $H$ by a random $q$-ary double circulant matrix.

*2) Quasi-dyadic construction:* We can also imagine a construction based on quasi-dyadic codes as proposed in [13].

Recently several *new* structural attacks appeared in [19] and [7] that extract the private key of some parameters of the variants presented in [2] and [13]. But in our context we deal with random codes and we are threaten by this kind of attacks.

Furthermore in [6] the authors describe a secure implementation of the Stern scheme using quasi-circulant codes. Our proposal inherits of the good properties of the original Stern scheme face to leakage of information as SPA and DPA attacks.

The parameters using quasi-cyclic or quasi-dyadic randoms codes are $q = 256, n = 128, k = 64, \mathsf{wt}(s) = 49$ this gives a public key of 512 bits and a private key of 1024 bits for almost the same complexity for an ISD attack.

*3) Embedding the secret in the matrix:* In fact it is possible to still decrease the sizes obtained in the previous subsection. The idea (from [10]) consists in embedding the secret key $x$ in the public matrix. To achieve that, we consider the secret as a word of the dual code of the code generated by the public matrix $H$. This means that we will use a null syndrom, which does not change the zero-knowledge property. We will detail this improvement in a longer version of this paper.

## V. Conclusion

We have defined an identification scheme which among all the schemes based on the SD problem has the best parameters for the size of the public data as well as for the communication complexity. Moreover, we have proposed a variant so as to reduce the size of the public data.

The improvement proposed here to the Stern scheme can be applied to all the Stern-based identification and signature schemes (as identity-based identification and signature [5] or threshold ring signature [12] for example).

We believe that this type of scheme is a realistic alternative to the usual number theory identification schemes in the case of constrained environments such as smart cards and of applications such as Pay-TV or vending machines.

### Acknowledgment

The authors want to thank Robert Niebuhr and Philippe Gaborit for their comments during the preparation of this paper and Christiane Peters for comments on her code.

## References

[1] S. Barg. Some new np-complete coding problems. *Probl. Peredachi Inf.*, 30:23–28, 1994.

[2] T. P. Berger, P.-L. Cayrel, P. Gaborit, and A. Otmani. Reducing key length of the McEliece cryptosystem. In *Progress in Cryptology – Africacrypt'2009*, Lecture Notes in Computer Science, pages 77–97. Springer, 2009.

[3] E. Berlekamp, R. McEliece, and H. van Tilborg. On the inherent intractability of certain coding problems. *IEEE Transactions on Information Theory*, 24(3):384–386, 1978.

[4] D. J. Bernstein, J. Buchmann, and E. Dahmen. *Post-Quantum Cryptography*. Springer, 2008.

[5] P.-L. Cayrel, P. Gaborit, and M. Girault. Identity-based identification and signature schemes using correcting codes. In *International Workshop on Coding and Cryptography, WCC 2007*, pages 69–78. editors : Augot, D., Sendrier, N., and Tillich, J.-P.

[6] P. L. Cayrel, P. Gaborit, and E. Prouff. Secure implementation of the stern authentication and signature schemes for low-resource devices. *CARDIS*, 2008.

[7] J.-C. Faugère, A. Otmani, L. Perret, and J.-P. Tillich. Algebraic cryptanalysis of McEliece variants with compact keys. 2009.

[8] A. Fiat and A. Shamir. How to prove yourself practical solutions to identification and signature problems. Springer LNCS 263:186–194, 1987.

[9] M. Finiasz and N. Sendrier. Security bounds for the design of code-based cryptosystems. In *to appear in Advances in Cryptology – Asiacrypt'2009*, 2009. http://eprint.iacr.org/2009/414.pdf.

[10] P. Gaborit and M. Girault. Lightweight code-based authentication and signature. In *IEEE International Symposium on Information Theory – ISIT'2007*, pages 191–195, Nice, France, 2007. IEEE.

[11] F. J. MacWilliams and N. J. A. Sloane. The theory of error correcting codes. In *North-Holland*, 1977.

[12] C. Aguilar Melchor, P.-L. Cayrel, and P. Gaborit. A new efficient threshold ring signature scheme based on coding theory. In *Post-quantum cryptography, second international workshop, PQCrypto 2008*, volume 5299 of *Lecture Notes in Computer Science*, pages 1–16. editors : J. Buchmann and J. Ding.

[13] R. Misoczki and P. S. L. M. Barreto. Compact mceliece keys from goppa codes. In *Selected Areas in Cryptography – SAC'09*, Proceedings, Calgary, Canada, 2009.

[14] R. Niebuhr and P. L. Cayrel. Generalizing information set decoding and the (generalized) birthday attack to $q$-ary codes. In *preprint*, 2009.

[15] C. Peters. Information-set decoding for linear codes over fq. In *Cryptology ePrint Archive, Report 2009/589*, 2009.

[16] J. N. Pierce. Limit distributions of the minimum distance of random linear codes. In *IEEE Trans. Inf. theory*, volume 13, pages 595–599, 1967.

[17] J. Stern. A method for finding codewords of small weight. In *Proc. of Coding Theory and Applications*, pages 106–113, 1989.

[18] J. Stern. A new identification scheme based on syndrome decoding. Lecture Notes in Computer Science vol. 773 Springer 1993:13–21, 1993.

[19] V. Gauthier Umana and G. Leander. Practical key recovery attacks on two McEliece variants. 2009. http://eprint.iacr.org/2009/509.pdf.

[20] P. Véron. Improved identification schemes based on error-correcting codes. *Appl. Algebra Eng. Commun. Comput.*, 8(1):57–69, 1996.